\documentclass[11pt,a4paper]{article}
\usepackage[hyperref]{acl2019}
\usepackage{times}
\usepackage{latexsym}
\usepackage{latexsym}
\usepackage{amsmath}
\usepackage{graphicx}
\usepackage{multicol}
\usepackage{multirow}
\usepackage{float}
\usepackage{caption}
\usepackage{mathtools}
\usepackage{courier}
\usepackage{qtree}
\usepackage{bbm}
\usepackage{array}
\usepackage{url}
\usepackage{color}
\usepackage{verbatim}
\usepackage{tabularx}

\usepackage{url}

\aclfinalcopy 

\title{A Just and Comprehensive Strategy for\\Using NLP to Address Online Abuse}

\author{David Jurgens \\
  University of Michigan \\
  School of Information \\
  \texttt{jurgens@umich.edu} \\
  \And
  Eshwar Chandrasekharan \\
  Georgia Tech \\
  School of Interactive Computing   \\
  \texttt{eshwar3@gatech.edu} \\
  \And
  Libby Hemphill \\
  University of Michigan \\
  School of Information \\  
  \texttt{libbyh@umich.edu} \\
}

\date{}

\begin{document}
\maketitle
\begin{abstract}

Online abusive behavior affects millions and the NLP community has attempted to mitigate this problem by developing technologies to detect abuse.  However, current methods have largely focused on a narrow definition of abuse to detriment of victims who seek both validation and solutions.  In this position paper, we argue that the community needs to make three substantive changes: (1) expanding our scope of problems to tackle both more subtle and more serious forms of abuse, (2) developing proactive technologies that counter or inhibit abuse before it harms, and (3) reframing our effort within a framework of justice to promote healthy communities.

\end{abstract}

\section{Introduction}

Online platforms have the potential to enable substantial, prolonged, and productive engagement for many people. Yet, the lived reality on social media platforms falls far short of this potential \cite{papacharissi2004democracy}. In particular, the promise of social media has been hindered by  antisocial, abusive behaviors such as harassment, hate speech, trolling, and the like.  Recent surveys indicate that abuse happens much more frequently than many people suspect (40\% of Internet users report being the subject of online abuse at some point), and members of underrepresented groups are targeted even more often~\cite{herring2002searching, drake2015darkest, Anti-Defamation_League2019-xf}.  

\begin{figure}[t]
    \centering
    \includegraphics[width=0.45\textwidth]{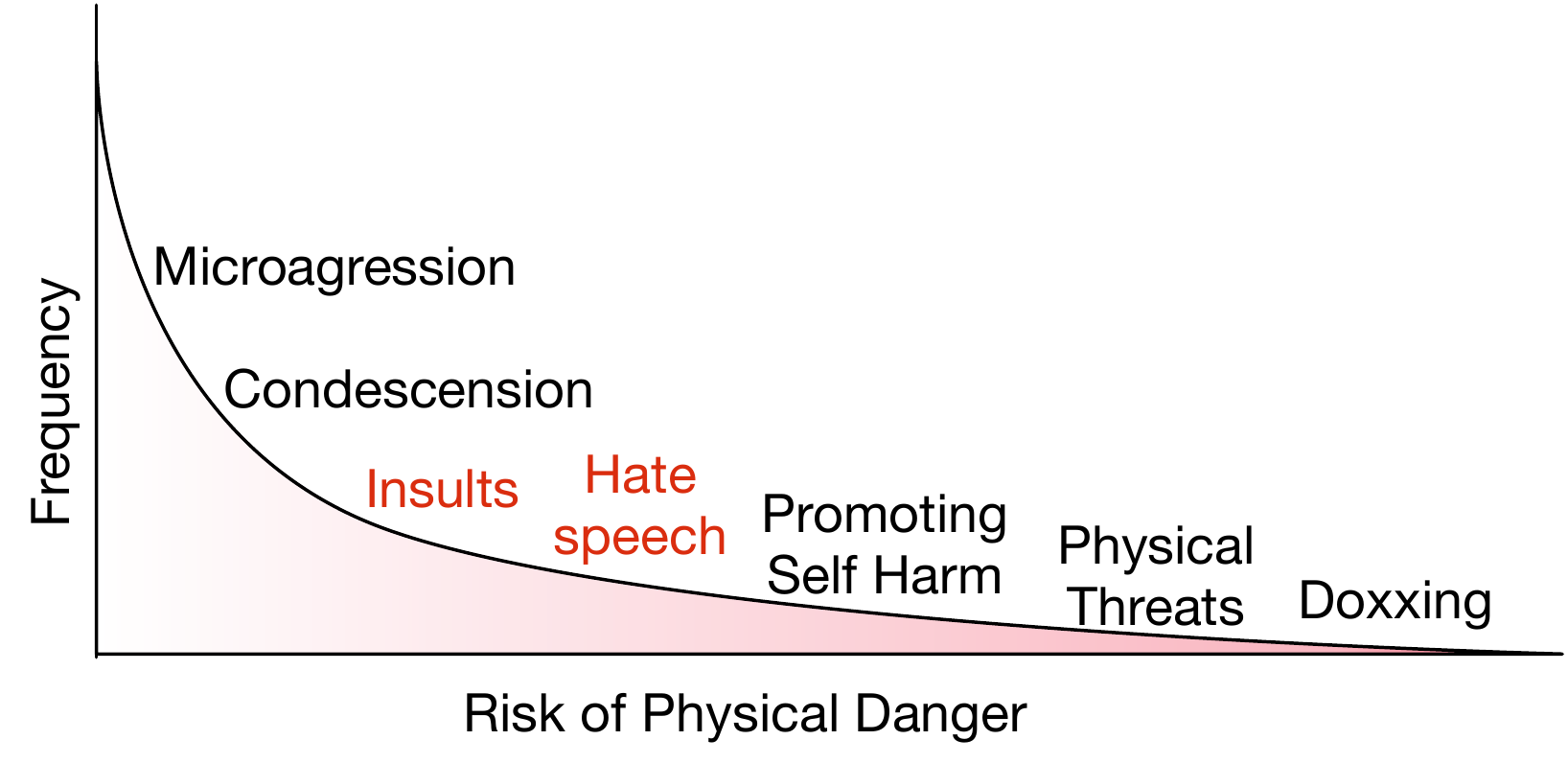}
    \caption{Abusive behavior online falls along a spectrum, and current approaches focus only on a narrow range (shown in red text), ignoring nearby problems. Impact comes from both the frequency (on left) and real-world consequences (on right) of behaviors. This figure illustrates the spectrum of online abuse in an hypothetical manner, with its non-exhaustive examples inspired from prior surveys of online experiences~\cite{duggan2017online, salminen2018anatomy}.}
    \label{fig:abuse-spectrum}
\end{figure}

The NLP community has responded by developing technologies to identify certain types of abuse and facilitating automatic or computer-assisted content moderation.  
Current technology has primarily focused on overt forms of abusive language and hate speech, without considering both (i) the success and failure of technology beyond getting the classification correct, and (ii) the myriad forms that abuse can take. As Figure \ref{fig:abuse-spectrum} shows, a large spectrum of abusive behavior exists---some with life-threatening consequences---much of which is currently unaddressed by language technologies. Explicitly hateful speech is just one tool of hate, and related tactics such as rape threats, gaslighting, First Amendment panic, and veiled insults are effectively employed both off- and online to silence, scare, and exclude participants from what should be inclusive, productive discussions \cite{Filipovic2007-sr}.

In this position paper, we argue that to promote healthy online communities, three changes are needed.  First, the NLP community needs to rethink and expand what constitutes abuse.  Second, current methods are almost entirely \textit{reactive} to abuse, entailing that harm occurs.  Instead, the community needs to develop proactive technologies that assist authors, moderators, and platform owners in preventing abuse before it occurs.  Finally, we argue that both of these threads point to a need for a broad re-aligning of our community goals towards \textit{justice}, rather than simply the elimination of abusive behavior.
In arguing for these changes, we outline how each effort offers new challenging NLP tasks that have concrete benefits.

\section{Rethinking What Constitutes Abuse}

The classifications we adopt and computationally enforce have real and lasting consequences by defining both what is and what is not abuse \cite{bowker2000sorting}.
Abusive behavior is an omnibus term that often includes harassment, threats, racial slurs, sexism, unexpected pornographic content, and insults---all of which can be directed at other users or at whole communities~\cite{Davidson2017-ua,Nobata2016-om}.
However, NLP has largely considered a far narrower scope of what constitutes abuse through its selection of which types of behavior to recognize~\cite{waseem2017understanding,schmidt2017survey,fortuna2018survey}. We argue that NLP needs to expand its computational efforts to recognize two additional general types of abuse: (a) infrequent and physically dangerous abuse,  and (b) more common but subtle abuse.  Additionally, we need to develop methods that respect community norms in classification decisions. These categories of abuse and the importance of community norms have been noted elsewhere \cite{liu2018forecasting,Guberman2017-sz,salminen2018anatomy,blackwell2017classification} but have not yet received the same level of attention in NLP.

Who has a right to speak and in what manner are subjective decisions that are guided by social relationships \cite{Foucault1972-bs, Noble2018-ip}, and the specific choices our algorithms make about what speech to allow and what to silence have powerful effects. 
For instance, rejecting behavior as not being abusive because it is outside the scope of our classification can cause substantial harm to victims \cite{blackwell2017classification}, tacitly involving the NLP community in algorithmic bias that sanctions certain forms of abuse.  Thus, categorization is particularly thorny: a broad categorization is likely too computationally inefficient, yet a narrow categorization risks further marginalizing affected community members and can lead to lasting harm.
Following, we outline three key directions for the community to expand its definitions.

\subsection{Physically Threatening Online Abuse}
We outline three computational challenges related to infrequent but overt physically-manifesting abuse that NLP could be applied to solve.
First, such behaviors do not necessarily adopt the language of hate speech or more common forms of hate speech and may in some contexts appear innocuous but are clearly dangerous in others.  For example, posting a phone number to call could be acceptable if one is encouraging others to call their political representative, yet would be a serious breach of privacy (doxxing) if posted as part of a public harassment campaign.  Similarly, declarations of ``keep up the weight loss!'' may be positive in a dieting community, yet reinforce dangerous behavior in a pro-anorexia community. Speech that in isolation appears offensive, such as impoliteness or racial slurs, may serve pro-social functions such as promoting intimacy \cite{Culpeper1996-tj} or showing camaraderie \cite{Allan2015-db}.  

Second, behaviors such as swatting, human trafficking, or pedophilia have all occurred on public social media platforms \cite{jaffe2016swatting,latonero2011human,holt2010considering}.  However, methods have yet to be developed for recognizing when users are engaging in these behaviors, which may involve coded language, and require  recognizing these alternative forms.  Current approaches for learning new explicitly-hateful symbols could be adapted to this task \citep[e.g.,][]{roy20161115cuck, gao2017recognizing}.
Third, online platforms have been used to incite mobs of people to violence \cite{siegel2015sectarian}.  These efforts often use incendiary fake news that plays upon factional rivalries \cite{samory2018conspiracies}.
Abusive  language detection methods can build upon recent advances at detecting fake news to identify content-sharing likely to lead to violence \cite{whatsappfakenews2018, oshikawa2018survey}.

\subsection{Subtle Abuse}\label{sec:subtle-abuse}
Many forms of abusive behavior are linguistically subtle and implicit.  Behaviors such as condescension, minimization (e.g., ``your situation isn't that bad''), benevolent stereotyping, and microagressions are frequently experienced by members of minority social groups \cite{Sue2007,glick2001ambivalent}. While subtle, such abuse can still be as emotionally harmful as overt abuse to some individuals \citep{Sue2010,Nadal2014}.  The NLP community has two clear paths for growth into this area.

First, although recognized within the larger NLP abuse typology \cite{waseem2017understanding}, only a handful of approaches have attempted these problems, such as identifying benevolent sexism \cite{jha2017does}, and new methods must be developed to identify the implicit 
signals.  Successful approaches will likely require advances in natural language understanding, as the abuse requires reasoning about the implications of the propositions.  A notable example of such an approach is 
\newcite{dinakar2012common} who extract implicit assumptions in statements and use common sense reasoning to identify social norm violations that would be considered insults.

Second, new methods should identify \textit{disparity} in treatment of social groups.  For example, in a study of the respectfulness of police language, \newcite{voigt2017language} found that officers were consistently less likely to use respectful language with black community members than with white community members---a disparity in a \textit{positive} social dimension. As NLP solutions have been developed for other social dimensions of language such as politeness \cite{danescu2013computational,munkova2013identifying,chhaya2018frustrated} and 
formality \cite{brooke2010automatic,sheikha2011generation,pavlick2016empirical}, these methods could be readily adapted for identifying such systematic bias for additional social categories and settings.

\subsection{Community Norms Need to be Respected}
Social norms are rules and standards that are understood by members of a group, and that guide and constrain social behavior without the force of laws~\cite{triandis1994culture,cialdini1998social}. 
Norms can be nested, in that they can be adopted from the general social context (e.g., use of pejorative adjectives are rude), and more general internet comment etiquette (e.g., using all caps is equivalent to shouting).
Yet, norms for what is considered acceptable can vary significantly from one community to another, making it challenging to build one abuse detection system that works for all communities~\cite{chandrasekharan2018internet}.

Current NLP methods are largely context- and norm-agnostic, which leads to situations where content is removed unnecessarily when deemed inappropriate (i.e., false positives), eroding community trust in the use of computational tools to assist in moderation.
A common failure mode for sociotechnical interventions like automated moderation is failing to understand the online community where they are being deployed~\cite{tumbler2018AI}.
Such community-specific norms and context are important to take into account, as NLP researchers are doubling down on context-sensitive approaches to define \citep[e.g.,][]{chandrasekharan2019hybrid} and detect abuse \citep[e.g.,][]{gao2017detecting}.

However, not all community norms are socially acceptable within the broader world.
Even behavior considered harmful in one community might be celebrated in another, e.g., Reddit's r/fatpeoplehate~\cite{chandrasekharan2017you}, and Something Awful Forums~\cite{pater2014just}. 
The existence of problematic normative behaviors within certain atypical online communities poses a challenge to abuse detection systems. \newcite{Fraser1990-lu} notes that when a public space is governed by a dominant group, its norms about participation end up perpetuating inequalities.
One approach to address this challenge would be to work closely with the different stakeholders involved in online governance, like platform administrators, policy makers, users and moderators. This will enable the development of solutions that cater to a wider range of expectations around moderating abusive behaviors on the platform, especially when dealing with deviant communities.

\subsection{Challenges for Creating New NLP Shared Tasks on Abusive Behavior}

Shared tasks have long been an NLP tradition for establishing evaluating metrics, defining data guidelines, and, more broadly, bringing together researchers.
The broad nature of abusive behavior creates significant challenges for the shared task paradigm. 
Here, we outline three opportunities for new shared tasks in this area.
First, new NLP shared tasks should develop annotation guidelines accurately define what constitutes abusive behavior in the target community.
Recent works have begun to make progress in this area by modeling the context in which a comment is made through user and community-level features~\cite{qian2018leveraging,mishra2018author,ribeiro2018characterizing}, yet often the norms in these settings are implicit making it difficult to transfer the techniques and models to other settings.
As one potential solution, \newcite{chandrasekharan2018internet} studied community norms on Reddit in a large-scale, data-driven manner, and released a dataset of over 40K removed comments from Reddit labeled according to the specific type of \textit{norm} being violated \cite{chandrasekharan2019hybrid}.

Second, new NLP shared tasks must address the data scarcity faced by abuse detection research while minimizing harm caused by the data. 
Constant exposure to abusive content has been found to negatively and substantially affect the mental health of moderators and users~\cite{roberts2014behind, gillespie2018custodians, saha2019prevalence}.
However, labeled ground truth data for building and evaluating classifiers is hard to obtain because platforms typically do not share moderated content due to privacy, ethical and public relations concerns.
One possibility for significant progress is to work with platform administrators and stakeholders to make proprietary data available as private test sets on platforms like Codalab, thereby keeping annotations in line with community norms and still allowing researchers to evaluate on real behavior.

Third, tasks must clearly define \textit{who} is the end-user of the classification labels.  For example, will moderators use the system to triage abusive content, or is the goal to automatically remove abusive content?
Current solutions are often trained and evaluated in a \textit{static} manner, only using preexisting data; whether these solutions are effective upon deployment remains relatively unexplored.
Evaluation must go beyond just traditional measures of performance like precision and recall, and instead begin optimizing for metrics like reduction in moderator effort, speed of response, targeted recall for severe types of abuse, moderator trust and fairness in predictions.

\section{Proactive Approaches for Abuse}

Existing computational approaches to handle abusive language are primarily reactive and intervene only after abuse has occurred. A complementary approach is developing \textit{proactive} technologies that prevent the harm from occurring in the first place, and we motivate three proactive computational approaches to prevent abuse here.

First, bystanders can have a profound effect on the course of an interaction by steering the direction of the conversation away from abuse \cite{markey2000bystander,dillon2015unresponsive}.  Prior work has used experimenter-based intervention but a substantial opportunity exists to operationalize these interventions through computational means.
\newcite{munger2017tweetment} developed a simple, but effective, computational intervention for the use of toxic language (the n-word), where a human-looking bot account would reply with a fixed comment about the harm such language caused and an appeal to empathy, leading to long-term behavior change in the offenders.  
Identifying how to best respond to abusive behavior---or whether to respond at all---are important computational next steps for this NLP strategy and one that likely needs to be done in collaboration with researchers from fields such as Psychology. 
Prior work has shown counter speech  to be effective for limiting the effects of hate speech~\cite{
schieb2016governing, mathew2018thou, stroud2018varieties}.
\newcite{wright2017vectors} notes that real-world examples of bystanders intervening can be found online, thereby providing a potential source of training data but methods are needed to reliably identify such counter speech examples.

Second, interventions that occur after a point of escalation may have little positive effect in some circumstances.  For example, when two individuals have already begun insulting one another, both have already become upset and must lose face to reconcile \cite{rubin1994social}.  At this point, de-escalation may prevent further abuse but does little for restoring the situation to a constructive dialog \cite{gottman1999marriage}.  However, interventions that occur \textit{before} the point of abuse can serve to shift the conversation.  Recent work has shown that it is possible to predict whether a conversation will become toxic on Wikipedia \cite{zhang2018conversations} and whether bullying will occur on Instagram \cite{liu2018forecasting}.  These predictable abuse trajectories open the door to developing new models for preemptive interventions that directly mitigate harm. 

Third, messages that are not intended as offensive create opportunities to nudge authors towards correcting their text if the offense is pointed out. This strategy builds upon recent work on explainable ML for identifying which parts of a message are offensive \cite{carton2018extractive,noever2018machine}, and work on paraphrase and style transfer for suggesting an appropriate inoffensive alternative \cite{santos2018fighting,prabhumoye2018style}.  For example, parts of a message could be paraphrased to adjust the level of politeness in order to minimize any cumulative disparity towards one social group \cite{sennrich2016controlling}.

\section{Justice Frameworks for NLP}

Martin Luther King Jr. wrote that the biggest obstacle to Black freedom is the ``white moderate, who is more devoted to `order' than to justice, who prefers a negative peace which is the absence of tension to a positive peace which is the presence of justice'' \cite{King1963-is}.  Analogously, by focusing only on classifying individual unacceptable speech acts, NLP risks being the same kind of obstacle as the white moderate: Instead of seeking the absence of certain types of speech, we should seek the presence of equitable participation. We argue that NLP should consider supporting three types of justice---social justice, restorative justice, and procedural justice---that describe (i) 
what actions are allowed and encouraged, 
(ii) how wrongdoing should be handled, and (iii) what procedures should be followed.

First, the \textit{capabilities approach to social justice} focuses on what actions people can do within a social setting \cite{Sen2011-ch,Nussbaum2003-ia} and provides a useful framework for thinking about what justice online could look like. \newcite{Nussbaum2003-ia} provides a set of 10 fundamental capabilities for a just society, such as the ability to express emotion and to have an affiliation.  
These capabilities provide a blueprint for articulating the values and opportunities an online community provides: Instead of a negative articulation---an ever-growing list of prohibited behaviors---we should use a positive phrasing (e.g., ``you will be able to'') of capabilities in an online community.  Such effort naturally extends our proposal for detecting community-specific abuse to one of promoting community norms.  Accordingly, NLP technologies can be developed to identify positive behaviors and ensure individuals are able to fulfill these capabilities.  
Several recent works have made strides in this direction by examining positive behaviors such as how constructive conversations are \cite{kolhatkar2017constructive,napoles2017finding}, whether dialog on contentious topics can exist without devolving into squabbling \cite{tan2016winning}, or the level of support given between community members \cite{wang2018s}.  

Second, once we have adequately articulated what people in a community should be able to do, we must address how the community handles transgressions. The notion of \textit{restorative justice} is a useful theoretical tool for thinking about how wrongdoing should be handled. Restorative justice theory emphasizes repair and uses a process in which stakeholders, including victims and transgressors, decide together on consequences. A restorative process may produce a punishment, such as banning, but can include consequences such as apology and reconciliation~\cite{Braithwaite2002-ca}. Just responses consider the emotions of both perpetrators and victims in designing the right response \cite{Sherman2003-tu}. A key problem here is identifying which community norm is violated and NLP technologies can be introduced to aid this process of elucidating violations through classification or use of explainable ML techniques.  Here, NLP can aid all parties (platforms, victims, and transgressors) in identifying appropriate avenues for restorative actions.

Third, just communities also require just means of addressing wrongdoing. The notion of \textit{procedural justice} explains that people are more likely to comply with a community's rules if they believe the authorities are legitimate \cite{Tyler2002-yj, Sherman2003-tu}.
For NLP, it means that our systems for detecting non-compliance must be transparent and fair. People will comply only if they accept the legitimacy of both the platform and the algorithms it employs. 
Therefore, abuse detection methods are needed to justify why a particular act was a violation to 
build legitimacy; a natural starting point for NLP in building legitimacy is recent work from explainable ML  \cite{ribeiro2016should,lei2016rationalizing,carton2018extractive}.

\section{Conclusion}

Abusive behavior online affects a substantial amount of the population.  The NLP community has proposed computational methods to help mitigate this problem, yet has also struggled to move beyond the most obvious tasks in abuse detection.  Here, we propose a new strategy for NLP to tackling online abuse in three ways.  First, expanding our purview for abuse detection to include both extreme behaviors and the more subtle---but still offensive---behaviors like microaggressions and condescension.
Second, NLP must develop methods that go beyond reactive identify-and-delete strategies to one of proactivity that intervenes or nudges individuals to discourage harm \textit{before it occurs}.  Third, the community should contextualize its effort inside a broader framework of justice---explicit capabilities, restorative justice, and procedural justice---to directly support the end goal of productive online communities.

\section*{Acknowledgements}
This material is based upon work supported by the Mozilla Research Grants program and by the National Science Foundation under Grant No. 1822228.

\bibliography{acl2019}

\begin{thebibliography}{84}
\expandafter\ifx\csname natexlab\endcsname\relax\def\natexlab#1{#1}\fi

\bibitem[{Allan(2015)}]{Allan2015-db}
Keith Allan. 2015.
\newblock When is a slur not a slur? the use of nigger in `pulp fiction'.
\newblock \emph{Lang. Sci.}, 52:187--199.

\bibitem[{{Anti-Defamation League}(2019)}]{Anti-Defamation_League2019-xf}
{Anti-Defamation League}. 2019.
\newblock Online hate and harassment: The american experience.
\newblock \url{https://www.adl.org/onlineharassment}.
\newblock Accessed: 2019-3-4.

\bibitem[{Blackwell et~al.(2017)Blackwell, Dimond, Schoenebeck, and
  Lampe}]{blackwell2017classification}
Lindsay Blackwell, Jill Dimond, Sarita Schoenebeck, and Cliff Lampe. 2017.
\newblock Classification and its consequences for online harassment: Design
  insights from heartmob.
\newblock \emph{Proceedings of the ACM on Human-Computer Interaction},
  1(CSCW):24.

\bibitem[{Bowker and Star(2000)}]{bowker2000sorting}
Geoffrey~C Bowker and Susan~Leigh Star. 2000.
\newblock \emph{Sorting things out: Classification and its consequences}.
\newblock MIT press.

\bibitem[{Braithwaite(2002)}]{Braithwaite2002-ca}
John Braithwaite. 2002.
\newblock \emph{Restorative Justice \& Responsive Regulation}.
\newblock Oxford University Press.

\bibitem[{Brooke et~al.(2010)Brooke, Wang, and Hirst}]{brooke2010automatic}
Julian Brooke, Tong Wang, and Graeme Hirst. 2010.
\newblock Automatic acquisition of lexical formality.
\newblock In \emph{Proceedings of the 23rd International Conference on
  Computational Linguistics: Posters}, pages 90--98. Association for
  Computational Linguistics.

\bibitem[{Carton et~al.(2018)Carton, Mei, and Resnick}]{carton2018extractive}
Samuel Carton, Qiaozhu Mei, and Paul Resnick. 2018.
\newblock Extractive adversarial networks: High-recall explanations for
  identifying personal attacks in social media posts.
\newblock In \emph{Proceedings of EMNLP}.

\bibitem[{Chandrasekharan and Gilbert(2019)}]{chandrasekharan2019hybrid}
Eshwar Chandrasekharan and Eric Gilbert. 2019.
\newblock Hybrid approaches to detect comments violating macro norms on reddit.
\newblock \emph{arXiv preprint arXiv:1904.03596}.

\bibitem[{Chandrasekharan et~al.(2017)Chandrasekharan, Pavalanathan,
  Srinivasan, Glynn, Eisenstein, and Gilbert}]{chandrasekharan2017you}
Eshwar Chandrasekharan, Umashanthi Pavalanathan, Anirudh Srinivasan, Adam
  Glynn, Jacob Eisenstein, and Eric Gilbert. 2017.
\newblock You can't stay here: The efficacy of reddit's 2015 ban examined
  through hate speech.
\newblock \emph{Proceedings of the ACM on Human-Computer Interaction},
  1(CSCW):31.

\bibitem[{Chandrasekharan et~al.(2018)Chandrasekharan, Samory, Jhaver, Charvat,
  Bruckman, Lampe, Eisenstein, and Gilbert}]{chandrasekharan2018internet}
Eshwar Chandrasekharan, Mattia Samory, Shagun Jhaver, Hunter Charvat, Amy
  Bruckman, Cliff Lampe, Jacob Eisenstein, and Eric Gilbert. 2018.
\newblock The internet's hidden rules: An empirical study of reddit norm
  violations at micro, meso, and macro scales.
\newblock \emph{Proceedings of the ACM on Human-Computer Interaction},
  2(CSCW):32.

\bibitem[{Chhaya et~al.(2018)Chhaya, Chawla, Goyal, Chanda, and
  Singh}]{chhaya2018frustrated}
Niyati Chhaya, Kushal Chawla, Tanya Goyal, Projjal Chanda, and Jaya Singh.
  2018.
\newblock Frustrated, polite, or formal: Quantifying feelings and tone in
  email.
\newblock In \emph{Proceedings of the Second Workshop on Computational Modeling
  of People’s Opinions, Personality, and Emotions in Social Media}, pages
  76--86.

\bibitem[{Cialdini and Trost(1998)}]{cialdini1998social}
Robert~B Cialdini and Melanie~R Trost. 1998.
\newblock Social influence: Social norms, conformity and compliance.
\newblock In D.~T. Gilbert, S.~T. Fiske, and G.~Lindzey, editors, \emph{The
  handbook of social psychology}, pages 151--192. McGraw-Hill.

\bibitem[{Culpeper(1996)}]{Culpeper1996-tj}
Jonathan Culpeper. 1996.
\newblock Towards an anatomy of impoliteness.
\newblock \emph{J. Pragmat.}, 25(3):349--367.

\bibitem[{Danescu-Niculescu-Mizil et~al.(2013)Danescu-Niculescu-Mizil, Sudhof,
  Jurafsky, Leskovec, and Potts}]{danescu2013computational}
Cristian Danescu-Niculescu-Mizil, Moritz Sudhof, Dan Jurafsky, Jure Leskovec,
  and Christopher Potts. 2013.
\newblock A computational approach to politeness with application to social
  factors.
\newblock In \emph{Proceedings of the Annual Meeting of the Association for
  Computational Linguistics (ACL)}.

\bibitem[{Davidson et~al.(2017)Davidson, Warmsley, Macy, and
  Weber}]{Davidson2017-ua}
Thomas Davidson, Dana Warmsley, Michael Macy, and Ingmar Weber. 2017.
\newblock Automated hate speech detection and the problem of offensive
  language.
\newblock In \emph{Eleventh International {AAAI} Conference on Web and Social
  Media}.

\bibitem[{Dillon and Bushman(2015)}]{dillon2015unresponsive}
Kelly~P Dillon and Brad~J Bushman. 2015.
\newblock Unresponsive or un-noticed?: Cyberbystander intervention in an
  experimental cyberbullying context.
\newblock \emph{Computers in Human Behavior}, 45:144--150.

\bibitem[{Dinakar et~al.(2012)Dinakar, Jones, Havasi, Lieberman, and
  Picard}]{dinakar2012common}
Karthik Dinakar, Birago Jones, Catherine Havasi, Henry Lieberman, and Rosalind
  Picard. 2012.
\newblock Common sense reasoning for detection, prevention, and mitigation of
  cyberbullying.
\newblock \emph{ACM Transactions on Interactive Intelligent Systems (TiiS)},
  2(3):18.

\bibitem[{Drake(2014)}]{drake2015darkest}
Bruce Drake. 2014.
\newblock The darkest side of online harassment: Menacing behavior.
\newblock \emph{Pew Research Center,
  http://www.pewresearch.org/fact-tank/2015/06/01/the-darkest-side-of-online-harassment-menacing-behavior/}.

\bibitem[{Duggan(2017)}]{duggan2017online}
Maeve Duggan. 2017.
\newblock Online harassment 2017.

\bibitem[{Filipovic(2007)}]{Filipovic2007-sr}
Jill Filipovic. 2007.
\newblock Blogging while female: How internet misogyny parallels
  ``{Real-World}'' harassment.
\newblock \emph{Yale J. Law Fem.}, 19(1).

\bibitem[{Fortuna and Nunes(2018)}]{fortuna2018survey}
Paula Fortuna and S{\'e}rgio Nunes. 2018.
\newblock A survey on automatic detection of hate speech in text.
\newblock \emph{ACM Computing Surveys (CSUR)}, 51(4):85.

\bibitem[{Foucault(1972)}]{Foucault1972-bs}
Michel Foucault. 1972.
\newblock \emph{The Archaeology of Knowledge \& The Discourse on Language}.
\newblock Pantheon Books, New York.

\bibitem[{Fraser(1990)}]{Fraser1990-lu}
Nancy Fraser. 1990.
\newblock Rethinking the public sphere: A contribution to the critique of
  actually existing democracy.
\newblock \emph{Social Text}, (25/26):56--80.

\bibitem[{Gao and Huang(2017)}]{gao2017detecting}
Lei Gao and Ruihong Huang. 2017.
\newblock Detecting online hate speech using context aware models.
\newblock In \emph{Proceedings of RANLP}.

\bibitem[{Gao et~al.(2017)Gao, Kuppersmith, and Huang}]{gao2017recognizing}
Lei Gao, Alexis Kuppersmith, and Ruihong Huang. 2017.
\newblock Recognizing explicit and implicit hate speech using a weakly
  supervised two-path bootstrapping approach.
\newblock In \emph{Proceedings of ICJNLP}.

\bibitem[{Gillespie(2018)}]{gillespie2018custodians}
Tarleton Gillespie. 2018.
\newblock \emph{Custodians of the Internet: Platforms, content moderation, and
  the hidden decisions that shape social media}.
\newblock Yale University Press.

\bibitem[{Glick and Fiske(2001)}]{glick2001ambivalent}
Peter Glick and Susan~T Fiske. 2001.
\newblock An ambivalent alliance: Hostile and benevolent sexism as
  complementary justifications for gender inequality.
\newblock \emph{American psychologist}, 56(2):109.

\bibitem[{Gottman(1999)}]{gottman1999marriage}
John~Mordechai Gottman. 1999.
\newblock \emph{The marriage clinic: A scientifically-based marital therapy}.
\newblock WW Norton \& Company.

\bibitem[{Guberman and Hemphill(2017)}]{Guberman2017-sz}
Joshua Guberman and Libby Hemphill. 2017.
\newblock Challenges in modifying existing scales for detecting harassment in
  individual tweets.
\newblock In \emph{Proceedings of the 50th Hawaii International Conference on
  System Sciences}.

\bibitem[{Herring et~al.(2002)Herring, Job-Sluder, Scheckler, and
  Barab}]{herring2002searching}
Susan Herring, Kirk Job-Sluder, Rebecca Scheckler, and Sasha Barab. 2002.
\newblock Searching for safety online: Managing" trolling" in a feminist forum.
\newblock \emph{The information society}, 18(5):371--384.

\bibitem[{Holt et~al.(2010)Holt, Blevins, and Burkert}]{holt2010considering}
Thomas~J Holt, Kristie~R Blevins, and Natasha Burkert. 2010.
\newblock Considering the pedophile subculture online.
\newblock \emph{Sexual Abuse}, 22(1):3--24.

\bibitem[{Jaffe(2016)}]{jaffe2016swatting}
Elizabeth~M Jaffe. 2016.
\newblock Swatting: the new cyberbullying frontier after elonis v. united
  states.
\newblock \emph{Drake L. Rev.}, 64:455.

\bibitem[{Jha and Mamidi(2017)}]{jha2017does}
Akshita Jha and Radhika Mamidi. 2017.
\newblock When does a compliment become sexist? analysis and classification of
  ambivalent sexism using twitter data.
\newblock In \emph{Proceedings of the second workshop on NLP and computational
  social science}, pages 7--16.

\bibitem[{King(1963)}]{King1963-is}
Martin~Luther King. 1963.
\newblock Letter from a birmingham jail.

\bibitem[{Kolhatkar and Taboada(2017)}]{kolhatkar2017constructive}
Varada Kolhatkar and Maite Taboada. 2017.
\newblock Constructive language in news comments.
\newblock In \emph{Proceedings of the First Workshop on Abusive Language
  Online}, pages 11--17.

\bibitem[{Krishna(2018)}]{tumbler2018AI}
Rachael Krishna. 2018.
\newblock Tumblr launched an algorithm to flag porn and so far it's just caused
  chaos, dec 2018.
\newblock
  \emph{\url{https://www.buzzfeednews.com/article/krishrach/tumblr-porn-algorithm-ban}}.

\bibitem[{Latonero(2011)}]{latonero2011human}
Mark Latonero. 2011.
\newblock Human trafficking online: The role of social networking sites and
  online classifieds.
\newblock \emph{Available at SSRN}.

\bibitem[{Lei et~al.(2016)Lei, Barzilay, and Jaakkola}]{lei2016rationalizing}
Tao Lei, Regina Barzilay, and Tommi Jaakkola. 2016.
\newblock Rationalizing neural predictions.
\newblock In \emph{Proceedings of EMNLP}.

\bibitem[{Liu et~al.(2018)Liu, Guberman, Hemphill, and
  Culotta}]{liu2018forecasting}
Ping Liu, Joshua Guberman, Libby Hemphill, and Aron Culotta. 2018.
\newblock Forecasting the presence and intensity of hostility on instagram
  using linguistic and social features.
\newblock In \emph{Twelfth International AAAI Conference on Web and Social
  Media}.

\bibitem[{Markey(2000)}]{markey2000bystander}
Patrick~M Markey. 2000.
\newblock Bystander intervention in computer-mediated communication.
\newblock \emph{Computers in Human Behavior}, 16(2):183--188.

\bibitem[{Mathew et~al.(2018)Mathew, Tharad, Rajgaria, Singhania, Maity, Goyal,
  and Mukherje}]{mathew2018thou}
Binny Mathew, Hardik Tharad, Subham Rajgaria, Prajwal Singhania, Suman~Kalyan
  Maity, Pawan Goyal, and Animesh Mukherje. 2018.
\newblock Thou shalt not hate: Countering online hate speech.
\newblock \emph{arXiv preprint arXiv:1808.04409}.

\bibitem[{McLaughlin(2018)}]{whatsappfakenews2018}
Timothy McLaughlin. 2018.
\newblock How whatsapp fuels fake news and violence in india.
\newblock
  \emph{https://www.wired.com/story/how-whatsapp-fuels-fake-news-and-violence-in-india/}.

\bibitem[{Mishra et~al.(2018)Mishra, Del~Tredici, Yannakoudakis, and
  Shutova}]{mishra2018author}
Pushkar Mishra, Marco Del~Tredici, Helen Yannakoudakis, and Ekaterina Shutova.
  2018.
\newblock \href {https://www.aclweb.org/anthology/C18-1093} {Author profiling
  for abuse detection}.
\newblock In \emph{Proceedings of the 27th International Conference on
  Computational Linguistics (COLING)}, pages 1088--1098.

\bibitem[{Munger(2017)}]{munger2017tweetment}
Kevin Munger. 2017.
\newblock Tweetment effects on the tweeted: Experimentally reducing racist
  harassment.
\newblock \emph{Political Behavior}, 39(3):629--649.

\bibitem[{Munkova et~al.(2013)Munkova, Munk, and
  Fr{\'a}terov{\'a}}]{munkova2013identifying}
Dasa Munkova, Michal Munk, and Zuzana Fr{\'a}terov{\'a}. 2013.
\newblock Identifying social and expressive factors in request texts using
  transaction/sequence model.
\newblock In \emph{Proceedings of RANLP}, pages 496--503.

\bibitem[{Nadal et~al.(2014)Nadal, Griffin, Wong, Hamit, and
  Rasmus}]{Nadal2014}
Kevin~L. Nadal, Katie~E. Griffin, Yinglee Wong, Sahran Hamit, and Morgan
  Rasmus. 2014.
\newblock \href {https://doi.org/10.1002/j.1556-6676.2014.00130.x} {{The impact
  of racial microaggressions on mental health: Counseling implications for
  clients of color}}.
\newblock \emph{Journal of Counseling and Development}, 92(1):57--66.

\bibitem[{Napoles et~al.(2017)Napoles, Tetreault, Pappu, Rosato, and
  Provenzale}]{napoles2017finding}
Courtney Napoles, Joel Tetreault, Aasish Pappu, Enrica Rosato, and Brian
  Provenzale. 2017.
\newblock Finding good conversations online: The yahoo news annotated comments
  corpus.
\newblock In \emph{Proceedings of the 11th Linguistic Annotation Workshop},
  pages 13--23.

\bibitem[{Nobata et~al.(2016)Nobata, Tetreault, Thomas, Mehdad, and
  Chang}]{Nobata2016-om}
Chikashi Nobata, Joel Tetreault, Achint Thomas, Yashar Mehdad, and Yi~Chang.
  2016.
\newblock Abusive language detection in online user content.
\newblock In \emph{Proceedings of the 25th International Conference on World
  Wide Web}, WWW '16, pages 145--153, Republic and Canton of Geneva,
  Switzerland. International World Wide Web Conferences Steering Committee.

\bibitem[{Noble(2018)}]{Noble2018-ip}
Safiya~Umoja Noble. 2018.
\newblock \emph{Algorithms of Oppression: How Search Engines Reinforce Racism}.
\newblock NYU Press.

\bibitem[{Noever(2018)}]{noever2018machine}
David Noever. 2018.
\newblock Machine learning suites for online toxicity detection.
\newblock \emph{arXiv preprint arXiv:1810.01869}.

\bibitem[{Nussbaum(2003)}]{Nussbaum2003-ia}
Martha Nussbaum. 2003.
\newblock Capabilities as fundamental entitlements: Sen and social justice.
\newblock \emph{Feminist Economics}, 9(2-3):33--59.

\bibitem[{Oshikawa et~al.(2018)Oshikawa, Qian, and Wang}]{oshikawa2018survey}
Ray Oshikawa, Jing Qian, and William~Yang Wang. 2018.
\newblock A survey on natural language processing for fake news detection.
\newblock \emph{arXiv preprint arXiv:1811.00770}.

\bibitem[{Papacharissi(2004)}]{papacharissi2004democracy}
Zizi Papacharissi. 2004.
\newblock Democracy online: Civility, politeness, and the democratic potential
  of online political discussion groups.
\newblock \emph{New media \& society}, 6(2):259--283.

\bibitem[{Pater et~al.(2014)Pater, Nadji, Mynatt, and Bruckman}]{pater2014just}
Jessica~Annette Pater, Yacin Nadji, Elizabeth~D Mynatt, and Amy~S Bruckman.
  2014.
\newblock Just awful enough: the functional dysfunction of the something awful
  forums.
\newblock In \emph{Proceedings of the 32nd annual ACM conference on Human
  factors in computing systems}, pages 2407--2410. ACM.

\bibitem[{Pavlick and Tetreault(2016)}]{pavlick2016empirical}
Ellie Pavlick and Joel Tetreault. 2016.
\newblock An empirical analysis of formality in online communication.
\newblock \emph{Transactions of the Association of Computational Linguistics
  (TACL)}, 4(1):61--74.

\bibitem[{Prabhumoye et~al.(2018)Prabhumoye, Tsvetkov, Salakhutdinov, and
  Black}]{prabhumoye2018style}
Shrimai Prabhumoye, Yulia Tsvetkov, Ruslan Salakhutdinov, and Alan~W Black.
  2018.
\newblock Style transfer through back-translation.
\newblock In \emph{Proceedings of ACL}.

\bibitem[{Qian et~al.(2018)Qian, ElSherief, Belding, and
  Wang}]{qian2018leveraging}
Jing Qian, Mai ElSherief, Elizabeth Belding, and William~Yang Wang. 2018.
\newblock \href {https://doi.org/10.18653/v1/N18-2019} {Leveraging intra-user
  and inter-user representation learning for automated hate speech detection}.
\newblock In \emph{Proceedings of the 2018 Conference of the North {A}merican
  Chapter of the Association for Computational Linguistics (NAACL)}, pages
  118--123.

\bibitem[{Ribeiro et~al.(2018)Ribeiro, Calais, Santos, Almeida, and
  Meira~Jr}]{ribeiro2018characterizing}
Manoel~Horta Ribeiro, Pedro~H Calais, Yuri~A Santos, Virg{\'\i}lio~AF Almeida,
  and Wagner Meira~Jr. 2018.
\newblock Characterizing and detecting hateful users on twitter.
\newblock In \emph{Twelfth International AAAI Conference on Web and Social
  Media}.

\bibitem[{Ribeiro et~al.(2016)Ribeiro, Singh, and Guestrin}]{ribeiro2016should}
Marco~Tulio Ribeiro, Sameer Singh, and Carlos Guestrin. 2016.
\newblock Why should i trust you?: Explaining the predictions of any
  classifier.
\newblock In \emph{Proceedings of KDD}, pages 1135--1144. ACM.

\bibitem[{Roberts(2014)}]{roberts2014behind}
Sarah~T Roberts. 2014.
\newblock \emph{Behind the screen: The hidden digital labor of commercial
  content moderation}.
\newblock Ph.D. thesis, University of Illinois at Urbana-Champaign.

\bibitem[{Roy(2016)}]{roy20161115cuck}
Jessica Roy. 2016.
\newblock "'cuck,''snowflake,''masculinist': A guide to the language of
  the'alt-right'.
\newblock
  \emph{\url{http://www.latimes.com/nation/la-na-pol-alt-right-terminology-20161115-story.html).
  Los Angeles Times}}.

\bibitem[{Rubin et~al.(1994)Rubin, Pruitt, and Kim}]{rubin1994social}
Jeffrey~Z Rubin, Dean~G Pruitt, and Sung~Hee Kim. 1994.
\newblock \emph{Social conflict: Escalation, stalemate, and settlement}.
\newblock Mcgraw-Hill Book Company.

\bibitem[{Saha et~al.(2019)Saha, Chandrasekharan, and
  De~Choudhury}]{saha2019prevalence}
Koustuv Saha, Eshwar Chandrasekharan, and Munmun De~Choudhury. 2019.
\newblock Prevalence and psychological effects of hateful speech in online
  college communities.
\newblock In \emph{WebSci}.

\bibitem[{Salminen et~al.(2018)Salminen, Almerekhi, Milenkovi{\'c}, Jung, An,
  Kwak, and Jansen}]{salminen2018anatomy}
Joni Salminen, Hind Almerekhi, Milica Milenkovi{\'c}, Soon-gyo Jung, Jisun An,
  Haewoon Kwak, and Bernard~J Jansen. 2018.
\newblock Anatomy of online hate: developing a taxonomy and machine learning
  models for identifying and classifying hate in online news media.
\newblock In \emph{Proceedings of the Twelfth International AAAI Conference on
  Web and Social Media (ICWSM)}.

\bibitem[{Samory and Mitra(2018)}]{samory2018conspiracies}
Mattia Samory and Tanushree Mitra. 2018.
\newblock Conspiracies online: User discussions in a conspiracy community
  following dramatic events.
\newblock In \emph{Twelfth International AAAI Conference on Web and Social
  Media}.

\bibitem[{Santos et~al.(2018)Santos, Melnyk, and Padhi}]{santos2018fighting}
Cicero Nogueira~dos Santos, Igor Melnyk, and Inkit Padhi. 2018.
\newblock Fighting offensive language on social media with unsupervised text
  style transfer.
\newblock In \emph{Proceedings of ACL}.

\bibitem[{Schieb and Preuss(2016)}]{schieb2016governing}
Carla Schieb and Mike Preuss. 2016.
\newblock Governing hate speech by means of counterspeech on facebook.
\newblock In \emph{Proceedings of ICA}, pages 1--23.

\bibitem[{Schmidt and Wiegand(2017)}]{schmidt2017survey}
Anna Schmidt and Michael Wiegand. 2017.
\newblock A survey on hate speech detection using natural language processing.
\newblock In \emph{Proceedings of the Fifth International Workshop on Natural
  Language Processing for Social Media}, pages 1--10.

\bibitem[{Sen(2011)}]{Sen2011-ch}
Amartya Sen. 2011.
\newblock \emph{The Idea of Justice}, reprint edition edition.
\newblock Belknap Press: An Imprint of Harvard University Press.

\bibitem[{Sennrich et~al.(2016)Sennrich, Haddow, and
  Birch}]{sennrich2016controlling}
Rico Sennrich, Barry Haddow, and Alexandra Birch. 2016.
\newblock Controlling politeness in neural machine translation via side
  constraints.
\newblock In \emph{Proceedings of NAACL}, pages 35--40.

\bibitem[{Sheikha and Inkpen(2011)}]{sheikha2011generation}
Fadi~Abu Sheikha and Diana Inkpen. 2011.
\newblock Generation of formal and informal sentences.
\newblock In \emph{Proceedings of the 13th European Workshop on Natural
  Language Generation}, pages 187--193. Association for Computational
  Linguistics.

\bibitem[{Sherman(2003)}]{Sherman2003-tu}
Lawrence~W Sherman. 2003.
\newblock Reason for emotion: Reinventing justice with theories, innovations,
  and research---the american society of criminology 2002 presidential address.
\newblock \emph{Criminology}, 41(1):1--38.

\bibitem[{Siegel(2015)}]{siegel2015sectarian}
Alexandra Siegel. 2015.
\newblock \emph{Sectarian Twitter Wars: Sunni-Shia Conflict and Cooperation in
  the Digital Age}, volume~20.
\newblock Carnegie Endowment for International Peace.

\bibitem[{Stroud and Cox(2018)}]{stroud2018varieties}
Scott~R Stroud and William Cox. 2018.
\newblock The varieties of feminist counterspeech in the misogynistic online
  world.
\newblock In \emph{Mediating Misogyny}, pages 293--310. Springer.

\bibitem[{Sue(2010)}]{Sue2010}
Derald~Wing Sue. 2010.
\newblock \emph{{Microaggressions in Everyday Life: Race, Gender, and Sexual
  Orientation}}.
\newblock Wiley, Hoboken, NJ.

\bibitem[{Sue et~al.(2007)Sue, Capodilupo, Torino, Bucceri, Holder, Nadal, and
  Esquilin}]{Sue2007}
Derald~Wing Sue, Christina~M Capodilupo, Gina~C Torino, Jennifer~M Bucceri,
  Aisha M.B.~B. Holder, Kevin~L Nadal, and Marta Esquilin. 2007.
\newblock \href {https://doi.org/10.1037/0003-066X.62.4.271} {{Racial
  microaggressions in everyday life: Implications for clinical practice}}.
\newblock \emph{American Psychologist}, 62(4):271--286.

\bibitem[{Tan et~al.(2016)Tan, Niculae, Danescu-Niculescu-Mizil, and
  Lee}]{tan2016winning}
Chenhao Tan, Vlad Niculae, Cristian Danescu-Niculescu-Mizil, and Lillian Lee.
  2016.
\newblock Winning arguments: Interaction dynamics and persuasion strategies in
  good-faith online discussions.
\newblock In \emph{Proceedings of the 25th international conference on world
  wide web}, pages 613--624. International World Wide Web Conferences Steering
  Committee.

\bibitem[{Triandis(1994)}]{triandis1994culture}
Harry~Charalambos Triandis. 1994.
\newblock \emph{Culture and social behavior}.
\newblock McGraw-Hill New York.

\bibitem[{Tyler and Huo(2002)}]{Tyler2002-yj}
Tom~R Tyler and Yuen Huo. 2002.
\newblock \emph{Trust in the Law: Encouraging Public Cooperation with the
  Police and Courts}.
\newblock Russell Sage Foundation.

\bibitem[{Voigt et~al.(2017)Voigt, Camp, Prabhakaran, Hamilton, Hetey,
  Griffiths, Jurgens, Jurafsky, and Eberhardt}]{voigt2017language}
Rob Voigt, Nicholas~P Camp, Vinodkumar Prabhakaran, William~L Hamilton,
  Rebecca~C Hetey, Camilla~M Griffiths, David Jurgens, Dan Jurafsky, and
  Jennifer~L Eberhardt. 2017.
\newblock Language from police body camera footage shows racial disparities in
  officer respect.
\newblock \emph{Proceedings of the National Academy of Sciences},
  114(25):6521--6526.

\bibitem[{Wang and Jurgens(2018)}]{wang2018s}
Zijian Wang and David Jurgens. 2018.
\newblock It's going to be okay: Measuring access to support in online
  communities.
\newblock In \emph{Proceedings of the 2018 Conference on Empirical Methods in
  Natural Language Processing}, pages 33--45.

\bibitem[{Waseem et~al.(2017)Waseem, Davidson, Warmsley, and
  Weber}]{waseem2017understanding}
Zeerak Waseem, Thomas Davidson, Dana Warmsley, and Ingmar Weber. 2017.
\newblock Understanding abuse: A typology of abusive language detection
  subtasks.
\newblock In \emph{Proceedings of the First Workshop on Abusive Language}.

\bibitem[{Wright et~al.(2017)Wright, Ruths, Dillon, Saleem, and
  Benesch}]{wright2017vectors}
Lucas Wright, Derek Ruths, Kelly~P Dillon, Haji~Mohammad Saleem, and Susan
  Benesch. 2017.
\newblock Vectors for counterspeech on twitter.
\newblock In \emph{Proceedings of the First Workshop on Abusive Language
  Online}, pages 57--62.

\bibitem[{Zhang et~al.(2018)Zhang, Chang, Danescu-Niculescu-Mizil, Dixon, Hua,
  Thain, and Taraborelli}]{zhang2018conversations}
Justine Zhang, Jonathan~P Chang, Cristian Danescu-Niculescu-Mizil, Lucas Dixon,
  Yiqing Hua, Nithum Thain, and Dario Taraborelli. 2018.
\newblock Conversations gone awry: Detecting early signs of conversational
  failure.
\newblock In \emph{Proceedings of ACL}.

\end{thebibliography}
\bibliographystyle{acl_natbib}

\end{document}